\documentclass{article}
\usepackage{spconf,amsmath,amssymb,graphicx}
\usepackage{enumitem}
\setlist{nosep, leftmargin=14pt}

\usepackage{mwe} 

\usepackage{xcolor}

\definecolor{bleucanard}{RGB}{0,139,139}

\title{Super-resolution of 4D flow MRI through inverse problem explicit solving}
\name{Aurélien de Turenne, Rémi Cart-Lamy and Denis Kouamé}
\address{IRIT, Université de Toulouse, CNRS, Toulouse INP, Toulouse, France}
\begin{document}
%
\maketitle
\begin{abstract}
Four-dimensional Flow MRI enables non-invasive, time-resolved imaging of blood flow in three spatial dimensions, offering valuable insights into complex hemodynamics. However, its clinical utility is limited by low spatial resolution and poor signal-to-noise ratio, imposed by acquisition time constraints. In this work, we propose a novel method for super-resolution and denoising of 4D Flow MRI based on the explicit solution of an inverse problem formulated in the complex domain. Using clinically available magnitude and phase images, we reconstruct synthetic complex-valued spatial signals. This enables us to model resolution degradation as a physically meaningful truncation of high-frequency components in k-space, and to recover high-resolution velocity fields through a fast, non-iterative 3D Fourier-based solver. The proposed approach enhances spatial resolution and reduces noise without the need for large training datasets or iterative optimization, and is validated on synthetic datasets generated from CFD simulations as well as on a 4D Flow MRI of a physical phantom.

\end{abstract}


\section{INTRODUCTION}
\label{sec:intro}  
Imaging of blood flow in clinical practice predominantly relies on two-dimensional phase-contrast MRI. More recently, four-dimensional phase-contrast MRI (also known as 4D Flow MRI) has emerged \cite{markl_4d_2012}. It provides time-resolved 3D velocity fields and enables detailed analysis of complex hemodynamics, for example in thoracic aortic diseases \cite{takahashi_utility_2022, adriaans_predictive_2019}. Despite its potential, the clinical adoption of 4D Flow MRI remains limited by low spatial and temporal resolution and suboptimal signal-to-noise ratio (SNR), imposed by acquisition time constraints. These limitations degrade the estimation of hemodynamic biomarkers such as wall shear stress \cite{levilly_towards_2020} and relative pressure fields \cite{krittian_finite-element_2012}.

To address these limitations, super-resolution methods have been proposed. Deep learning approaches \cite{fathi_super-resolution_2020, ferdian_4dflownet_2020} demonstrate strong performance but rely heavily on simulated training data, as acquiring paired low- and high-resolution in vivo datasets is generally infeasible. Consequently, their performance often degrades when applied to real clinical acquisitions, where noise characteristics and artefacts differ from those seen during training. Alternatively, inverse problem formulations leverage physical priors \cite{levilly_segmentation-free_2024}, yet they often involve computationally intensive iterative optimization and require careful parameter tuning. 

In this work, we introduce a fast, fully explicit method for 4D Flow MRI super-resolution and denoising based on the analytical solution of a linear inverse problem in the complex domain. Synthetic 3D complex-valued signals are reconstructed from clinically available magnitude and phase images, resolution loss is modeled as a k-space truncation, and high-resolution 3D velocity-encoded signals are recovered with a non-iterative 3D Fourier-domain solver adapted from \cite{zhao_fast_2016, tuador_novel_2021}. This explicit formulation avoids iterative refinement and requires tuning only one parameter, making the approach computationally efficient and simple to deploy in practice. The phase is extracted only after reconstruction, yielding super-resolved velocity fields without the need for training data.

Section~2 presents the acquisition process, forward model, and inversion method; Section~3 reports results on synthetic and phantom datasets; and Section~4 concludes the study and outlines perspectives toward clinical translation.

\section{METHODS}
\label{sec:methods}  
In this section, we describe the acquisition process of 4D Flow MRI and introduce the forward model. We then present our inverse problem formulation and the corresponding solution for super-resolving velocity imaging.

\subsection{Acquisition process}

In 4D Flow MRI, blood flow is measured by applying velocity-encoding bipolar gradients along three orthogonal directions. For each encoding direction, two complex-valued acquisitions are performed: a reference acquisition without flow encoding, and a second acquisition in which a bipolar gradient induces a velocity-dependent phase shift. The corresponding complex signals can be written as
\begin{equation}
s_{\text{ref}} = A\, e^{i \Phi_{\text{ref}}},
\end{equation}
\begin{equation}
\begin{aligned}
s_{\text{venc},j} &= A\, e^{i \left( \Phi_{\text{ref}} + \Phi_{j} \right)},
\end{aligned}
\quad j \in \{u,v,w\},
\end{equation}
where $A$ is the magnitude image, $\Phi_{\text{ref}}$ is a background phase term, and $\Phi_{j}$ is the velocity-induced phase shift along the $x$, $y$, or $z$ direction.

The velocity-encoding phase is obtained as the phase difference between the two acquisitions,
\begin{equation}
\Phi_{j}
= (\Phi_{\text{ref}}+\Phi_{j}) - \Phi_{\text{ref}}
, \quad j \in \{u,v,w\},
\end{equation}
and is linearly related to the corresponding velocity component $v_j$ through the VENC (Velocity ENCoding) parameter, which determines the maximum measurable velocity without aliasing:
\begin{equation}
\Phi_{j} = \frac{\pi}{\text{VENC}} \, v_j, \quad j \in \{u,v,w\}.
\end{equation}

\subsection{Forward model}
Although the scanner acquires complex-valued data for $s_{\text{ref}}$ and $s_{\text{venc},j}$ for $j \in \{u,v,w\}$, the DICOM output only contains:
\begin{itemize}
    \item one 3D magnitude image \(A\),
    \item three 3D phase images $\Phi_u$, $\Phi_v$ and $\Phi_w$.
\end{itemize}

Given the magnitude and phase images, we reconstruct for each encoding direction a synthetic 3D complex-valued spatial signal:
\begin{equation}
y_j = A\, e^{i \Phi_j}, \quad j \in \{u, v, w\}.
\end{equation}

Using these reconstructed 3D complex-valued signals $y_j$, we define a forward model that captures the loss of spatial resolution in clinical 4D Flow MRI acquisitions. Limited spatial resolution can be modeled as the truncation of high-frequency components in he k-space due to acquisition time constraints. In the spatial domain, this degradation can be modeled as a convolution with a sinc kernel followed by a subsampling. For each velocity-encoding direction, the acquisition model is written as:
\begin{equation}
y_j = S H x_j + n_j, \quad j \in \{u, v, w\},
\end{equation}
where $y_j \in \mathbb{C}^{N_l}$ ($N_l = m_l \times n_l \times s_l$) and $x_j \in \mathbb{C}^{N_h}$ ($N_h = m_h \times n_h \times s_h$) are respectively vectorized versions of LR image of size $m_l \times n_l \times s_l$ and HR image of size $m_h \times n_h \times s_h$, obtained by ordering their voxels lexicographically. The 3D HR image is altered by a decimation operator $\mathbf{S} \in \mathbb{R}^{N_l \times N_h}$ with an integer rate $d = d_r \times d_c \times d_s$, i.e., $N_h = N_l \times d$. $\mathbf{H} \in \mathbb{R}^{N_h \times N_h}$ is a convolution operator modeled as a BCCB matrix of the kernel, and $\mathbf{n} \in \mathbb{R}^{N_l \times 1}$ is an additive white Gaussian noise. The decimation rates $d_r$, $d_c$ and $d_s$ correspond to the pixel resolution loss in each spatial direction, satisfying $m_h = m_l \times d_r$, $n_h = n_l \times d_c$ and $s_h = s_l \times d_s$. 

This formulation allows us to model super-resolution as an inverse problem on the complex-valued signal. Once a high-resolution estimate $\hat{x}_j \in \mathbb{C}^{N_h}$ is recovered, we extract its phase component and convert it to velocity using the VENC relation. The final output of the pipeline is thus a super-resolved velocity field in each direction, which is the clinically relevant quantity for flow analysis.

\subsection{Inverse problem solving}
For each encoding direction $j \in \{u,v,w\}$ and for each time step $t$, we seek a high-resolution estimate $x_j$ of the 3D complex-valued velocity-encoded signal, from which the final velocity field will be recovered by phase extraction and conversion via the VENC relation. From the forward model, we formulate the following Tikhonov-regularized inverse problem:
\begin{equation}
\min_{x_j} \ \frac{1}{2} \| y_j - S H x_j \|_2^2 + \tau \| x_j - \bar{x}_j \|_2^2, 
\quad j \in \{u, v, w\},
\end{equation}

where $\bar{x}_j \in \mathbb{C}^{N_h}$ is a rough (interpolated) estimate of the high-resolution signal and $\tau$ is a regularization parameter controlling the trade-off between data fidelity and regularization. 

The solutions can be expressed as:
\begin{equation}
\begin{aligned}
\hat{x}_j &= \left( H^H S^H S H + 2\tau I_{N_h} \right)^{-1} 
             \left( H^H S^H y_j + 2\tau \bar{x}_j \right), \\
&\quad j \in \{u, v, w\},
\end{aligned}
\end{equation}
where $.^H$ represents the Hermitian transpose. Direct inversion of the matrix \((H^H S^H S H + 2\tau I_{N_h})\) is not computationally feasible due to the high dimensionality of \(H\). To avoid iterative optimization methods such as the Alternating Direction Method of Multipliers (ADMM), we rely on the FSR (Fast Super-Resolution) approach proposed by Zhao et al.~\cite{zhao_fast_2016}, and later extended to 3D by Tuador et al.~\cite{tuador_novel_2021}. Unlike these methods, which apply super-resolution to real-valued images, we operate directly in the complex domain and extract the phase images only after the reconstruction. This choice reflects the fact that the forward degradation model is linear with respect to the complex signal \(A e^{i\Phi}\), whereas the extraction of the phase \(\Phi\) is a non-linear operation.

Based on the FSR approach, the closed-form solution for each velocity encoding direction can be written as:
\begin{equation}
\begin{aligned}
\hat{x}_j &= \frac{1}{2\lambda}k_j 
- \frac{1}{2\lambda} F^H \underline{\Lambda}^H 
\left( 2\lambda \tau I_{N_l} + \underline{\Lambda} \underline{\Lambda}^H \right)^{-1} 
\underline{\Lambda} Fk_j, \\
&\quad j \in \{u, v, w\},
\end{aligned}
\end{equation}
where 
\begin{equation}
k_j = H^H S^H y_j + 2\tau \bar{x}_j,
\end{equation}
and
\begin{equation}
\underline{\Lambda} = 
\left( 
\left( \mathbf{1}_{d_s}^T \otimes I_{s_l} \right)
\otimes 
\left( \mathbf{1}_{d_c}^T \otimes I_{n_l} \right)
\otimes 
\left( \mathbf{1}_{d_r}^T \otimes I_{m_l} \right)
\right)
\Lambda,
\end{equation}
with $\mathbf{1}_{u}^T \in \mathbb{R}^{1 \times u}$ a row vector of ones, 
$I_v \in \mathbb{R}^{v \times v}$ the identity matrix, and $\otimes$ the Kronecker product.

This closed-form expression avoids direct inversion of large matrices and only requires one 3D FFT and one inverse FFT, combined with pointwise operations in the frequency domain. It thus enables fast, non-iterative super-resolution of the complex-valued velocity-encoded signals.

\section{RESULTS}
\label{sec:results}
In this section, we evaluate the proposed method on both synthetic and physical phantom data. Both ×2 and ×4 super-resolution factors were tested, with similar overall trends. For clarity, we show the most illustrative case for each dataset: ×4 for simulated data and ×2 for the phantom experiment.

\subsection{Datasets}
The proposed method was evaluated on two datasets:  
(i) synthetic data generated from computational fluid dynamics (CFD) simulations of three aortic geometries from the work of Ferdian et al.~\cite{ferdian_4dflownet_2020}, and  
(ii) a 4D Flow MRI acquisition of a physical flow phantom from Zimmermann et al.~\cite{zimmermann_impact_2021}. For the synthetic dataset, low-resolution (LR) inputs were created by truncating high-frequency k-space components and adding complex Gaussian noise (PSNR~15~dB). High-resolution (HR) references correspond directly to the CFD-generated velocity fields.  
For the phantom dataset, LR images were obtained from the clinical acquisition, while no HR reference is available.

\subsection{Simulated dataset results}
\begin{figure}[htbp]
    \centering
    \includegraphics[width=0.5\textwidth]{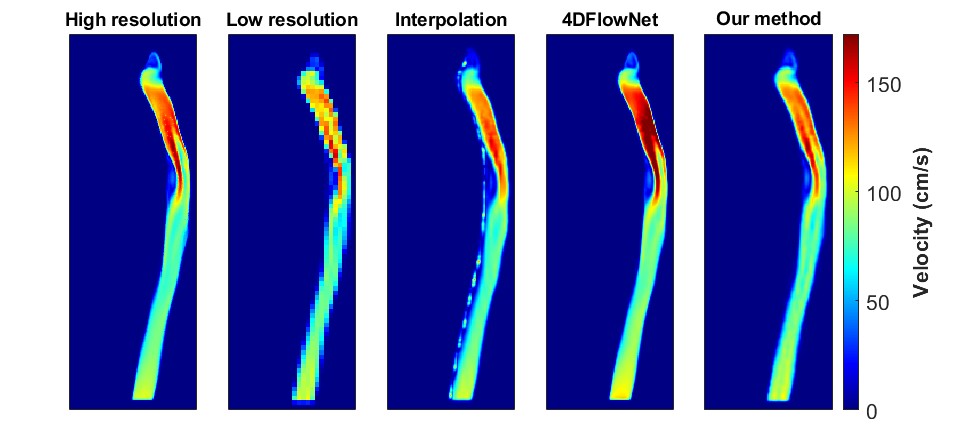}
    \caption{Comparison between the HR reference, LR input, bicubic interpolation, 4DFlowNet, and our method for a $\times 4$ super-resolution task on synthetic data. The velocity magnitude (cm/s) is shown within the flow mask.}
    \label{fig:qualitative}
\end{figure}

Fig.~\ref{fig:qualitative} shows a visual comparison for the $\times 4$ super-resolution task on the synthetic data. Bicubic interpolation excessively smooths the velocity field and removes fine-scale flow structures. 4DFlowNet produces noise free smoother patterns but tends to overestimate velocity values. In contrast, our method provides a more balanced reconstruction, preserving small-scale structures while avoiding velocity overestimation.

\begin{figure}[htbp]
    \centering
    \includegraphics[width=0.45\textwidth]{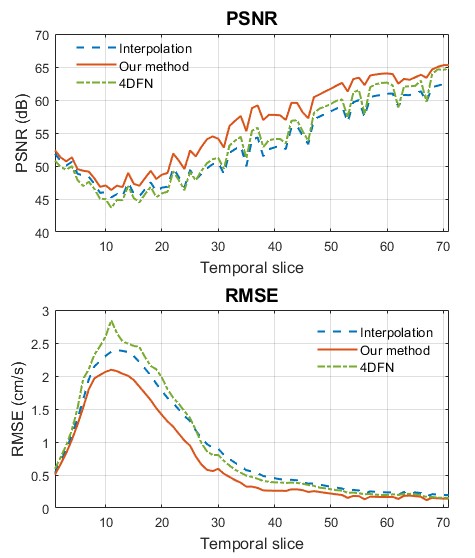}
    \caption{Quantitative evaluation on synthetic data ($\times 4$). PSNR and RMSE computed within the flow mask over all time frames. Our method achieves the best reconstruction fidelity.}
    \label{fig:metrics}
\end{figure}

To quantitatively assess reconstruction quality, we computed two metrics inside the flow mask for each time frame and for all three velocity components ($u$, $v$, $w$):
\begin{itemize}
    \item \textbf{PSNR} between reconstructed and reference velocity components,
    \item \textbf{Root Mean Square Error (RMSE)} between reconstructed and reference components.
\end{itemize}
Fig.~\ref{fig:metrics} summarizes the results for the $\times 4$ task.  
The proposed method achieves the best performance across both metrics. It provides higher PSNR values, indicating a more faithful reconstruction of the velocity components, and lower RMSE, reflecting improved preservation of velocity magnitudes. Bicubic interpolation systematically underperforms due to its strong smoothing effect, while 4DFlowNet shows increased errors related to velocity overestimation.  
These results confirm that explicitly solving the inverse problem yields more accurate and temporally stable reconstructions.

\subsection{Phantom dataset results}
\begin{figure}[htbp]
    \centering
    \includegraphics[width=0.5\textwidth]{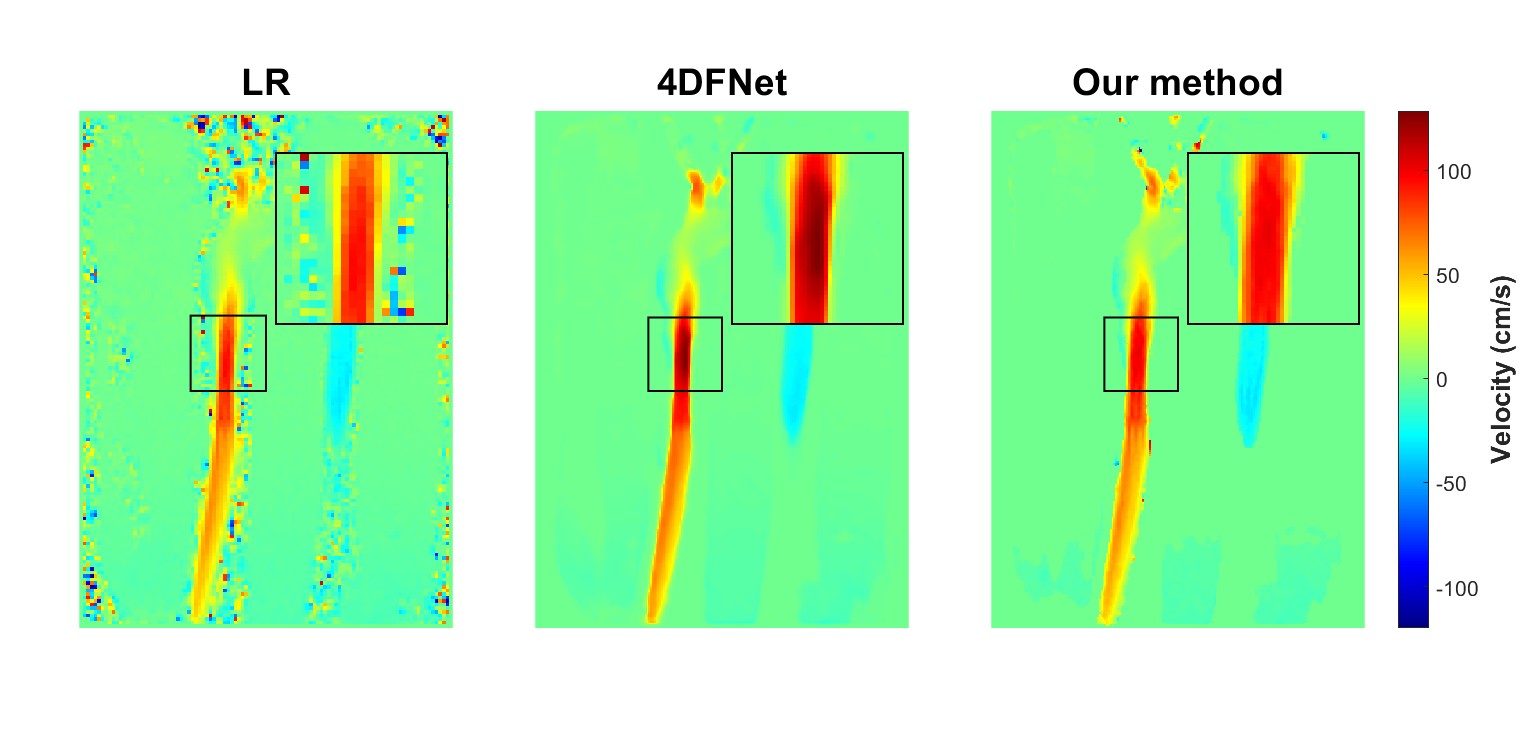}
    \caption{Comparison between the LR input, 4DFlowNet, and our method on a physical flow phantom for a $\times 2$ super-resolution task. The in-plane velocity component (cm/s) is shown within the flow mask.}
    \label{fig:phantom}
\end{figure}

Fig.~\ref{fig:phantom} provides visual results for the physical phantom in the $\times 2$ setting. The LR acquisition exhibits noise and blurred velocity patterns. 4DFlowNet produces a smoother image but tends to overestimate the velocity magnitude. In contrast, the proposed method yields a more consistent reconstruction, maintaining realistic velocity values while reducing noise and preserving coherent flow structures.

\section{Conclusion}
We introduced a fast, explicit method for super-resolution and denoising of 4D Flow MRI based on the analytical solution of a complex-domain inverse problem. By reconstructing synthetic 3D complex-valued signals from standard magnitude and phase images and using a non-iterative Fourier-based solver, the method enhances spatial resolution while remaining computationally efficient and training-free.

The approach currently relies on Tikhonov regularization, and its validation has been limited to CFD-based simulations and a physical phantom. Future work will investigate the application of the proposed method to patient data, in order to assess its performance in capturing spatial–temporal flow structures and its ability to generalize to complex in vivo conditions. The integration of more advanced priors will also be explored to further improve reconstruction quality and robustness. Hybrid approaches such as deep unfolding will be investigated to automate parameter selection and enhance robustness while preserving the interpretability of the proposed framework.

\bibliographystyle{IEEEbib}
\bibliography{refs}

\end{document}